\documentclass[prb,twocolumn,floats,aps]{revtex4}
\usepackage{graphicx,graphics,color,epsfig,multirow} 
\usepackage{bm}
\usepackage{amssymb}
\usepackage{amsmath}
\usepackage{amsfonts}

\newcommand{\ket}[1]{\left| #1 \right>}

\begin{document}

\title{Local integrals of motion in the two-site Anderson-Hubbard model}

\author{R. Wortis$^{1}$ and Malcolm P. Kennett$^{2}$}
\affiliation{$^1$Department of Physics \& Astronomy, Trent University, 1600 West Bank Dr., Peterborough ON, Canada, K9J 7B8 \\
$^2$Department of Physics, Simon Fraser University, Burnaby, British Columbia, Canada, V5A 1S6}
\date{\today}

\date{\today}

\begin{abstract}
It has been proposed that the states of fully many-body localized systems can be described in terms of conserved local pseudospins. 
Due to the multitude of ways to define these, the explicit identification of the optimally local pseudospins in specific systems is non-trivial. 
Given continuing intense interest in the role of disorder in strongly correlated systems, we consider the disordered Hubbard model.
Focusing on a two-site system, we track the evolution of the optimally localized pseudospins as hopping and interactions are varied to move the system away from the trivially localized atomic limit, examining the explicit form of the pseudospins and exploring the broad distribution of non-optimal forms.
\end{abstract}


\maketitle

\section{Introduction}
\label{sec-intro}

In 1958, Anderson showed that disorder can cause the single-particle states in non-interacting systems to be localized in space.\cite{Anderson1958}
Spurred by the work of Basko {\it et al.},\cite{Basko2006} 
recent attention has focused on the persistence of this localization in the presence of interactions 
and the nature of the resulting many-body localized phase.\cite{Nandkishore2015,Altman2015,Huse2016}
In isolated quantum systems, key signatures of this phase include a lack of ergodicity\cite{Pal2010,Huse2014} 
and a logarithmic growth of entanglement entropy at long times.\cite{Bardason2012}
Several recent experiments in cold atom and trapped ion systems show evidence of many-body localized states.\cite{Schrieber2015,Kondov2015,Bordia2016,Choi2016,Smith2016}

The properties of fully many-body localized systems can be understood as arising from the presence of a macroscopic number of local conserved quantities.\cite{Vosk2013,Serbyn2013,Huse2014,
Chandran2015,Bera2015,Ros2015,Imbrie2016prl,Imbrie2016jsp,Rademaker2016,Pekker2016,Imbrie2016arx}
Most, although not all,\cite{Serbyn2013,Chandran2015} identifications of these local integrals of motion (LIOM) have associated them with pseudospins. \cite{Vosk2013,Huse2014,Ros2015,Pekker2016}
The Hamiltonian may be written in terms of these pseudospins as\cite{Huse2014}
\begin{eqnarray}
	{\cal H} &=& \sum_i \alpha_i \tau_i^z + \sum_{ij} \beta_{ij} \tau_i^z \tau_j^z + \sum_{ijk} \gamma_{ijk} \tau_i^z \tau_j^z \tau_k^z + \ldots ,
	\nonumber \\
	\label{eq:lbit_Hamiltonian}
\end{eqnarray}
where $\tau_i^z$ is the $z$-component of the $i^{\rm th}$ pseudospin.  
While this is true for any system for which Hilbert-space dimension is a power of two,\cite{Lychkovskiy2013}
a defining feature of the many-body localized phase is that these pseudospins can be chosen to be local.
Indeed choice is involved -- LIOM are not uniquely defined, because combinations of conserved quantities are themselves conserved.   
In non-interacting Anderson localized systems, it is convenient to identify the LIOM 
as the occupancies of the localized single-particle states such that the Hamiltonian is expressed in the form of Eq.~(\ref{eq:lbit_Hamiltonian}) with only the $\alpha_i$ coefficients non-zero.\cite{Ros2015}  
In contrast, a MBL insulator will have interactions between the local pseudospins.  
A variety of different schemes have been proposed for mapping MBL systems onto a pseudospin Hamiltonian.\cite{Vosk2013,Serbyn2013,Huse2014,
Chandran2015,Bera2015,Ros2015,Imbrie2016prl,Imbrie2016jsp,Rademaker2016,Pekker2016,Monthus2016,Wahl2016,He2016,Inglis2016,OBrien2016,Rademaker2016arxiv,Bera2016}  

Much of the theoretical work on MBL has been for either spin systems\cite{Pal2010,Vosk2013,Serbyn2013,Huse2014,Chandran2015,Pekker2016} 
or equivalently spinless fermions.\cite{Bera2015,Ros2015,Rademaker2016}  
The many fascinating properties of doped transition metal oxides and of related cold atomic gas systems,
which are uniquely associated with the interplay of strong correlations and disorder,
motivate extending the study of MBL to Hubbard-type systems.
In particular, while there has been some theoretical\cite{Mondaini2015,BarLev2016,Znidaric2016} 
and experimental\cite{Kondov2015} studies of disordered Hubbard models, 
the focus has not been on identifying the LIOM.

Here we address two questions.
First, what form do the integrals of motion take in the Anderson-Hubbard model?
Second, how can we identify the most local choice of these integrals of motion?
In addressing these questions, 
we hope to obtain 
insight into the real space interpretation of localization in a many-body system.

To address these questions in a transparent manner, we consider the toy system of the two-site Anderson-Hubbard model.
The two-site Anderson,\cite{Johri2012} Hubbard,\cite{Avella2003,Ziesche1997,Murmann2014,Balcerzak2016} 
and Anderson-Hubbard model\cite{Wortis2010,Chen2010,Perera2015} have provided insights into the physics of these respective models.
We take advantage of the simplicity of this system to explore all possible choices of pseudospins for which the single-particle eigenstates of the system are connected to the vacuum by a single pseudospin raising operator.
We search the tremendous multiplicity of options to identify the most local choice, using a localization measure based on the strength of the support of the pseudospin operator on a 
single site and focusing on minimizing not the average but the largest localization measure in the set of pseudospins.
We present maps of the strength of localization as a function of the on-site interaction strength $U$ and the hopping amplitude $t$.
We show explicit expressions for the resulting pseudospin operators in terms of Fock space 
creation and annihilation operators, 
and demonstrate the wide distribution of localization strengths
which arise for different choices of pseudospin mapping.

This paper is structured as follows: 
In Sec.~\ref{sec:model} we introduce the two-site Anderson-Hubbard model, discuss the choices available in 
identifying pseudospins
and present our method of searching for the most local option.
In Sec.~\ref{sec:results} we present the results we obtain from numerical optimization of the locality of the LIOMs 
and in Sec.~\ref{sec:discussion} discuss the implications of our results.

\section{Model and approach}
\label{sec:model}

In this section we introduce the two-site Anderson-Hubbard model and our approach to obtain optimally localized LIOMs.  
For context we briefly present natural choices of integrals of motion in two simple limits (no hopping and no interactions),
before considering in detail the optimization of the locality of the LIOMs in the general case. 

\subsection{Two-site Anderson-Hubbard model}

The Anderson-Hubbard model is a tight-binding model which combines the on-site 
Coulomb repulsion of the Hubbard model with the disorder of the Anderson model.  For a two-site system,
the Hamiltonian takes the form
\begin{eqnarray}
{\cal H} &=& -t \sum_{\sigma=\uparrow,\downarrow} \left( {\hat c}_{1\sigma}^{\dag} {\hat c}_{2\sigma} + {\hat c}_{2\sigma}^{\dag} {\hat c}_{1\sigma} \right) \nonumber \\
& & + \, U \sum_{i=1,2} {\hat n}_{i\uparrow} {\hat n}_{i\downarrow} + \sum_{i=1,2, \sigma=\uparrow,\downarrow} \epsilon_i {\hat n}_{i\sigma}, 
\label{eq:twositeHAmodel}
\end{eqnarray}
where ${\hat c}_{i\sigma}^{\dag}$ ($\hat{c}_{i\sigma}$) is the creation (annihilation) operator for an electron 
with spin $\sigma$ at lattice site $i$ and ${\hat n}_{i\sigma}={\hat c}_{i\sigma}^{\dag} {\hat c}_{i\sigma}$ is the number operator for spin $\sigma$ at site $i$.  
The site potentials $\epsilon_i$ are randomly chosen from a 
probability distribution such as a Gaussian or a uniform distribution with fixed width.
In the present study our focus is on the properties of individual systems as opposed to disorder-averaged quantities, and 
we therefore let $\epsilon_1 = 0.5$ and $\epsilon_2 = -0.5$ such that the difference in potential between the two sites 
$|\epsilon_1 - \epsilon_2|$ sets 
the energy scale for the hopping amplitude $t$ and the interaction strength $U$.

This model is sufficiently simple that analytic expressions for all eigenvalues and eigenvectors can be found, 
as summarized in Appendix\ \ref{app:2siteAHM}.  The dimension of the Hilbert space is 
$2^4=16$
which implies that the Hamiltonian can be mapped to four pseudospins.
Our main focus here is exploring the representation of the model in terms of LIOMs, a task for which the analytic solutions are convenient but not required.  

\subsection{Identifying LIOMs} 
In this section we discuss the construction of LIOMs for the two-site Anderson-Hubbard model,
and the optimization of the locality of these LIOMs.  
Before tackling the general case, we start with a discussion 
of the pseudospin representation of the model in two simple limits in which one can immediately identify integrals of motion: the atomic limit ($t=0$, $U \neq 0$) 
and the non-interacting limit ($t\neq 0$, $U=0$).

\subsubsection{The atomic limit:  $t=0$, $U \neq 0$}
\label{sec:tequals0}
The simplest limit is when hopping $t=0$ and the system is trivially localized.
The Hamiltonian contains only number operators
${\hat n}_{i\sigma} = {\hat c}_{i\sigma}^{\dag} {\hat c}_{i\sigma}$ 
and we can 
write down pseudospins by inspection.  The first can be related to fermion creation and annihilation
operators via
\begin{equation} 
\tau_1^+ \equiv {\hat c}_{1\uparrow}^{\dag}, \quad \quad  
\tau_1^- = (\tau_1^+)^{\dag} = {\hat c}_{1\uparrow}, 
\end{equation}
\begin{equation} 
\tau_1^z = \tau_1^+ \tau_1^- - \frac{1}{2} = n_{1\uparrow} - \frac{1}{2}.
\end{equation}
Similarly we may define $\tau_2^+={\hat c}_{2\uparrow}^{\dag}$, $\tau_3^+={\hat c}_{1\downarrow}^{\dag}$ and 
$\tau_4^+={\hat c}_{2\downarrow}^{\dag}$, which gives
\begin{eqnarray}
{\cal H}_{t=0} &=& \epsilon_1 (\tau_1^z + \tau_3^z) + \epsilon_2 (\tau_2^z + \tau_4^z) \nonumber \\
& & + \, U \tau_1^z \tau_3^z + U \tau_2^z \tau_4^z + {\rm constants}.
\end{eqnarray}
In this case the integrals of motion are simply the numbers of particles of each spin at each site, $n_{i\sigma}$, up to a constant.
These are conserved and maximally local.

\subsubsection{The non-interacting limit: $t \neq 0$,  $U=0$}
\label{sec:Uequals0}
A second case which provides useful context is the non-interacting limit (i.e. $U = 0$).  
Here we know that the many-body states can be expressed in terms of the 
occupancies of a set of single-particle states, which for the two-site system are just the bonding and anti-bonding orbitals with corresponding creation operators 
\begin{eqnarray}
{\hat a}_{+\sigma}^\dagger &=& \alpha {\hat c}_{1\sigma}^{\dag} - \beta {\hat c}_{2\sigma}^{\dag} , \\
{\hat a}_{-\sigma}^\dagger &=& \beta {\hat c}_{1\sigma}^{\dag} + \alpha {\hat c}_{2\sigma}^{\dag}.
\end{eqnarray}
The coefficients $\alpha$ and $\beta$ as well as the corresponding energies $E_{1p}$ and $E_{1m}$ can be determined by diagonalizing 
Eq.~(\ref{eq:twositeHAmodel}) with $U = 0$ and are listed in Appendix~\ref{app:2siteAHM}.
If we let 
\begin{eqnarray}
\tau_1^+={\hat a}^\dagger_{+\uparrow}, \ \tau_2^+={\hat a}^\dagger_{-\uparrow}, \ \tau_3^+={\hat a}^\dagger_{+\downarrow} \ {\rm and } \ \tau_4^+={\hat a}^\dagger_{-\downarrow}, 
\label{U0taus}
\end{eqnarray}
then the Hamiltonian can be written as
\begin{eqnarray}
{\cal H} &=& E_{1p} (\tau_1^z + \tau_2^z) + E_{1m} (\tau_3^z + \tau_4^z) + \, {\rm constants} .
\end{eqnarray}
These integrals of motion are (up to a constant) the number 
of particles of each spin in each orbital.
There is nothing inherently local about this set of integrals of motion.
When $t\ll |\epsilon_1-\epsilon_2|$ most of the weight for a given $\tau$ will be on a single site,
but in the limit $t\rightarrow \infty$ the states are as delocalized as they can be, having 
equal weight on the two sites.

When both $t$ and $U$ are non-zero, the choice of $\tau$ operators is less clear, and 
we address both how to define them in general and how to choose them to ensure that they are as local as possible.

\subsubsection{The general case: $t \neq 0$, $U \neq 0$}
\label{sec:taus_general}
In the cases considered in Secs.~\ref{sec:tequals0} and \ref{sec:Uequals0} the many-body eigenstates either correspond to occupied sites ($t=0$) or occupied bonding/anti-bonding orbitals ($U=0$).  
The pseudospin raising operators $\tau_i^+$ may thus be chosen to equal the corresponding fermion
creation and annihilation operators.  
When both hopping and interactions are present, the single-particle eigenstates are the same as in the non-interacting case, and so the raising operators chosen in Eq.~(\ref{U0taus}) connect the vacuum to the single-particle eigenstates.
However, applying two such raising operators will {\em not} result in the correct interacting two-particle eigenstates.
This implies that pseudospin raising operators need to be modified by including terms which have 
no effect on the vacuum, but have non-trivial effects on occupied states.
Given the many possible options, a more systematic approach is required than that used in Secs.~\ref{sec:tequals0} and \ref{sec:Uequals0}.

\begin{table}[htbp]
   \centering
\begin{tabular}{|l|l |r |} \hline
& eigenstate & $\tau$ state \\ \hline
& 0 				& $|\rangle_{\tau}$ \\ \hline
\multirow{4}{*}{\rotatebox[origin=c]{90}{1-particle}}
& $p\uparrow$  		& $\tau_1^+ |\rangle_{\tau}$ \\
& $m\uparrow$  	& $\tau_2^+ |\rangle_{\tau}$ \\
& $p\downarrow$  	& $\tau_3^+ |\rangle_{\tau}$ \\
& $m\downarrow$  	& $\tau_4^+ |\rangle_{\tau}$ \\ \hline
\multirow{6}{*}{\rotatebox[origin=r]{90}{2-particle}} 
& $t\uparrow$ 		& $\tau_2^+ \tau_1^+ |\rangle_{\tau}$ \\
& $t_0$ 			& $\tau_4^+ \tau_1^+ |\rangle_{\tau}$ \\
& $t\downarrow$  	& $\tau_4^+ \tau_3^+ |\rangle_{\tau}$ \\ 
& $w_1$  			& $\tau_3^+ \tau_1^+ |\rangle_{\tau}$ \\
& $w_2$  			& $-\tau_3^+ \tau_2^+ |\rangle_{\tau}$ \\
& $w_3$  			& $\tau_4^+ \tau_2^+ |\rangle_{\tau}$ \\ \hline
\multirow{6}{*}{\rotatebox[origin=l]{90}{3-particle}} 
& $3p\uparrow$ 	& $-\tau_3^+ \tau_2^+ \tau_1^+ |\rangle_{\tau}$ \\
& $3m\uparrow$ 	& $\tau_4^+ \tau_2^+ \tau_1^+ |\rangle_{\tau}$ \\
& $3p\downarrow$ 	& $\tau_4^+ \tau_3^+ \tau_1^+ |\rangle_{\tau}$ \\
& $3m\downarrow$ 	& $-\tau_4^+ \tau_3^+ \tau_2^+ |\rangle_{\tau}$ \\ \hline
& 4 				& $ -\tau_4^+ \tau_2^+ \tau_3^+ \tau_1^+ |\rangle_{\tau}$ \\ \hline
\end{tabular}
   \caption{The sixteen eigenstates (details in Appendix\ \ref{app:2siteAHM}) grouped by particle number and one possible way to match these with the $\tau$ states obtained by acting on the vacuum $|\rangle_{\tau}$ with raising operators $\tau_i^+$, referred to in the text as match 1.}
   \label{table:match1}
\end{table}

The eigenstates of the Hamiltonian are also eigenstates of the $\tau_i^z$ operators.
This informs our approach to identifying pseudospins, which relies on having all the energy eigenstates and proceeds in five steps.
First, list the eigenstates of the $\tau_i^z$ operators. 
These are obtained by acting on the vacuum with all possible combinations of the 
$\tau_i^+$ operators, as shown in the right column of Table~\ref{table:match1}.  
These states, including their signs and the order in which they are listed, define what we call the $\tau$ basis.

Second, match each of these states to one of the eigenstates of the system.  
A particular choice of match, match 1, is shown in Table~\ref{table:match1}.
In this choice, $n$-particle states are connected to the vacuum by $n$ $\tau_i^+$ operators.  
We emphasize that there is no requirement that such an identification is made.
Instead, for example, the four-particle state could be connected to the vacuum by a single $\tau$ operator if one so chose.  
The correspondence in Table~\ref{table:match1} was chosen to align with the pseudospin choice for the non-interacting case discussed above.

Third, write a particular pseudospin operator,
e.g. $\tau_1^+$ (which we use below to illustrate the approach), in the $\tau$ basis.  
Fourth, perform the unitary transformation from the $\tau$ basis to the 
Fock basis using the matrix of eigenvectors, $Q$, that diagonalizes the Hamiltonian in the Fock basis.
When the eigenstates are ordered as in Table~\ref{table:match1} the matrix $Q$ takes the form
\begin{eqnarray}
\underline{\underline Q} &=& 
\left( \begin{array} {c cc cc cc c} 
1 & 0 & 0 & 0 & 0 & 0 & 0 & 0 \\
		0 & \underline{\underline {Q_1}} & 0 & 0 & 0 & 0 & 0 & 0 \\
		0 & 0 & \underline{\underline {Q_1}} & 0 & 0 & 0 & 0 & 0 \\
		0 & 0 & 0 & \underline{\underline {Q_{2t}}} & 0 & 0 & 0 & 0 \\
		0 & 0 & 0 & 0 & \underline{\underline {Q_{2s}}} & 0 & 0 & 0 \\
		0 & 0 & 0 & 0 & 0 & \underline{\underline {Q_3}} & 0 & 0 \\
		0 & 0 & 0 & 0 & 0 & 0 & \underline{\underline {Q_3}} & 0 \\
0 & 0 & 0 & 0 & 0 & 0 & 0 & 1 \\
\end{array}
\right) , \\
\underline{\underline {Q_1}} &=& \left( \begin{array} {cc} \alpha & \beta \\ - \beta & \alpha \end{array} \right) , 
\quad \quad \underline{\underline {Q_3}} \ = \ \left( \begin{array} {cc} \alpha & -\beta \\ \beta & \alpha \end{array} \right) = 
\underline{\underline {Q_1}}^T ,\\
\underline{\underline {Q_{2t}}} &=& \left( \begin{array} {ccc} 1 & 0 & 0 \\ 0 & 1 & 0 \\ 0 & 0 & 1 \end{array} \right) , \quad 
\quad \underline{\underline {Q_{2s}}} \ = \ \left( \begin{array} {ccc} \phi_{1\ell} & \phi_{2\ell} & \phi_{3\ell} \\
\phi_{1s} & \phi_{2s} & \phi_{3s}  \\
\phi_{1r} & \phi_{2r} & \phi_{3r} 
 \end{array} \right)  , \nonumber \\ & &
\end{eqnarray}
where the $\phi$s are defined in Appendix~\ref{app:2siteAHM}.
Hence we may write, e.g. $\tau_1^+$ in the Fock basis as
\begin{eqnarray}
\left.\tau_1^+\right|_{\rm Fock \ basis} &=& Q \left. \tau_1^+\right|_{\tau \ {\rm basis}} Q^{\dag} .
\end{eqnarray}

Fifth, project the $\tau_1^+$ operator, written in the Fock basis, onto all possible combinations of ${\hat c}_{i\sigma}$ and ${\hat c}_{i\sigma}^{\dag}$ operators.  These include the identity operator, the single operators ${\hat c}_{1\uparrow}$ and ${\hat c}_{1\uparrow}^{\dag}$, etc., and all possible combinations of two, three and four annihilation and creation operators, up to and including the four fermion operator ${\hat n}_{1\uparrow} {\hat n}_{2\uparrow} {\hat n}_{1\downarrow} {\hat n}_{2\downarrow}$.
There are 256 unique combinations, consistent with the number of independent quantities in a 16$\times$16 matrix.
One can construct matrix representations of all of these operators, starting from the eight 
single $c$ operators in the Fock basis.  
We choose this matrix representation to be orthonormal in the following sense:
\begin{eqnarray}
\sum_{i,j=1}^{16} A_{ij} B_{ij} &=& \biggl\{ \begin{array}{ll} 1 & {\rm for} \ A=B \\ 0 & {\rm for} \ A \ne B \end{array}   .
\end{eqnarray}
For terms involving the number operators orthogonality is achieved by working with   
${\underline{\underline{{\tilde n}_{i\sigma}}}} \equiv \underline{\underline{n_{i\sigma}}} - \frac{1}{2} \underline{\underline{I}}$.

There is an additional subtlety in the atomic limit.
When $t=0$, the triplet state $\ket{t_0} = (\ket{\uparrow \downarrow} + \ket{\downarrow \uparrow})/\sqrt{2}$ is degenerate with the singlet state $\ket{s} = (\ket{\uparrow \downarrow} - \ket{\downarrow \uparrow})/\sqrt{2}$.
In this case any arbitrary orthogonal superposition of $\ket{t_0}$ or $\ket{s}$ may be assigned as $\tau$ states.
In the results we present below, when $t=0$ we identify $\ket{\uparrow \downarrow}$ and $\ket{\downarrow \uparrow}$ as eigenstates rather than $\ket{t_0}$ and $\ket{s}$.  This ensures maximally localized $\tau$ operators in this case.
For $t\neq 0$ this degeneracy is absent and no special treatment is required.

The principle of this method, namely that the pseudospins can be identified using the unitary transformation between the pseudospin basis and a convenient product-state basis, has been raised by a number of authors\cite{Huse2014, Chandran2015, Pekker2016} but not directly implemented.
Moreover, to our knowledge the only other explicit attempt to optimize the choice of LIOMs is in Ref.~\onlinecite{He2016} for a spin model as opposed to a Hubbard model.

\subsubsection{Maximizing the locality of LIOMs}
In Sec.~\ref{sec:taus_general} we presented the identification of pseudospins, and in Appendix\ \ref{app:H_tauz} we
write the Hamiltonian in terms of their $z$-components.
The claim for many-body localization is that this identification can be 
made such that the pseudospins are local.\cite{Huse2014}
As emphasised above there is no unique identification of the pseudospins. 
In the two-site Anderson-Hubbard model, at first glance, there are at least 16! possible matchings between energy eigenstates and $\tau$ states.  
To find the most local choice, we must search through these.
While daunting in systems where the Hilbert space is large,\cite{Chandran2015} this can be done in the small system we consider.
Even so, in order to reduce the amount of computation required, we make use of several simplifications.

First, we focus only on $\tau_i^+$ operators which correspond to the 
addition of a single particle, so that $n$-particle states are connected to the vacuum state by $n$ pseudospin raising operators.  
Second, since the labelling of the $\tau_i^+$ operators themselves is arbitrary, 
we consider only one of the correspondence options in the one-particle block.
We also note that the triplet state $\ket{t\uparrow}=\ket{\uparrow\uparrow}$ must be 
created by two spin-up $\tau^+$ operators and $\ket{t\downarrow}$ by two spin-down operators.
This leaves four possibilities for assigning $\tau$ operators to two-particle states.
An additional consideration which expands the number of options is the issue of signs.  
In assigning the correspondences for the two-particle states $\ket{t_0}$, $\ket{w_1}$, $\ket{w_2}$ and 
$\ket{w_3}$, we allow for matches with either a positive or a negative sign, and likewise for the three- and four-particle states.

With these considerations, there are 
$8\cdot6\cdot4\cdot2=384$ options in the two-particle block, 
384 in the three-particle block, and 2 in the four-particle block, 
for a total of 294,912 possible matches.
We now specify the criterion that we optimize in our search.

\subsubsection{Localization measure}

In a two-site system any definition of a localization length is somewhat artificial, but
clearly the system we study is trivially localized when $t=0$, and less localized for
non-zero $t$.  In the absence of interactions, if one starts at the atomic limit and
then gradually turns on hopping, one expects the single-particle states to evolve
so that there is weight on both sites.  The ratio of the weights on the two sites can 
be used to determine a measure of localization.  
In an interacting system, there are two ways in which a pseudospin operator can become nonlocal.
As in non-interacting systems, there can be a shift of weight from single-site operators on one site to single-site operators on another site, 
and in addition there can be a shift of weight to operators which act on multiple sites.
We use a simple measure which treats both of these possibilities on equal footing.  
Consider a specific $\tau_i^+$ operator from a specific match $m$ between $\tau$ states and eigenstates.
Let $w_1$ be the sum of the squares of the coefficients of all $c$ operators which act only on site 1 (e.g. $c_{1\uparrow}^{\dag}$, $c_{1\downarrow}$, etc).  
Let $w_2$ be the same quantity for site 2.
And finally let $w_b$ be the sum of the squares of the coefficients of all $c$ operators which act on {\em both} sites (e.g. $c_{1\uparrow}^{\dag} c_{2\uparrow}$, ${\tilde n}_{1\uparrow} c_{2\downarrow}$, etc).
$w_1+w_2+w_b=1$.  

We define
\begin{eqnarray}
\xi_{mi1} \ = \ {1-w_1 \over w_1}, \quad {\rm and} \quad \xi_{mi2} \ = \ {1 - w_2 \over w_2}, 
\end{eqnarray}
from which we obtain
\begin{eqnarray}
\xi_{mi} &=& {\rm min}(\xi_{mi1},\xi_{mi2}) .
\end{eqnarray}
$\xi_{mi}$ is a measure of locality for the $i^{\rm th}$ pseudospin in match $m$.
When all the weight is on site 1, $\xi_{mi}=\xi_{mi1}=0$, and likewise for site 2.  
When weight shifts off the primary site, $\xi_{mi}$ becomes nonzero.  
Only values $\xi_{mi} \ll 1$ can be considered localized.
In particular, if $w_b=0$ and $w_1=w_2$, $\xi_{mi}=1$.
Meanwhile, as $w_b \rightarrow 1$, $\xi_{mi} \rightarrow \infty$.

For a specific choice of the match between eigenstates and $\tau$ states,
there will be a $\xi_{mi}$ corresponding to each $\tau_i^+$.  
We characterize the match by $\xi_m={\rm max}_i\{ \xi_{mi} \}$, the maximum $\xi_{mi}$ value.  We then search over all matches to find the most local set of pseudospins, which we characterize with 
\begin{eqnarray}
\xi = {\rm min}_m \{\xi_{m}\}.
\label{loc_meas}
\end{eqnarray}

\section{Results}
\label{sec:results}
In this section we present first our numerical results for the evolution of the localization measure $\xi$ with the parameters $t$ and $U$ in the two-site Anderson-Hubbard model.
We then show the explicit form of some of the resulting LIOMs, and finally we display 
the distribution of $\xi_{mi}$ values obtained from the many possible matches explored.

\begin{figure}[htbp] 
   \centering
   \begin{tabular}{l}
   (a) \\
   \includegraphics[width=\columnwidth]{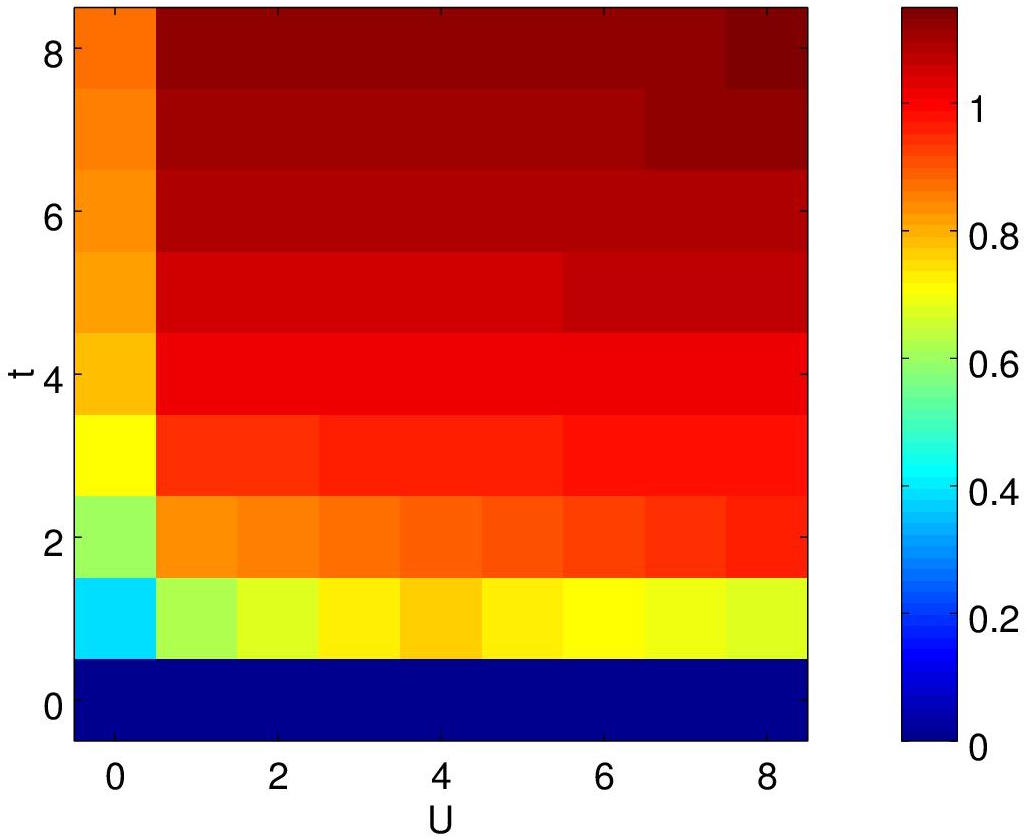} \\
   \begin{tabular}{cc}
   (b) & (c) \\
   \includegraphics[width=0.5\columnwidth]{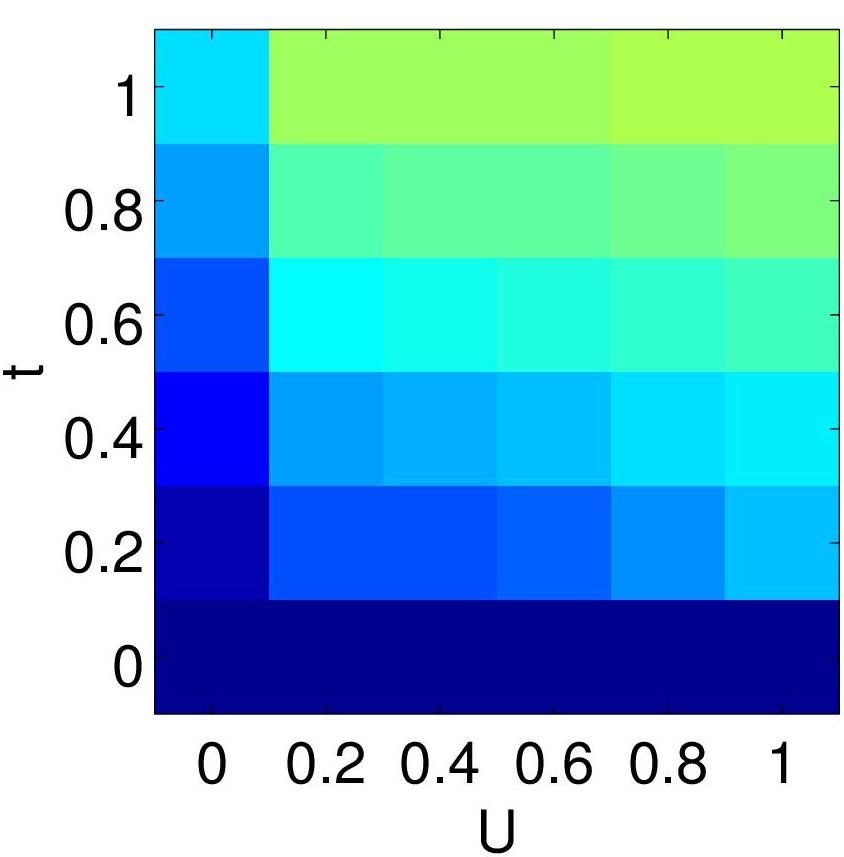}
   &
   \includegraphics[width=0.5\columnwidth]{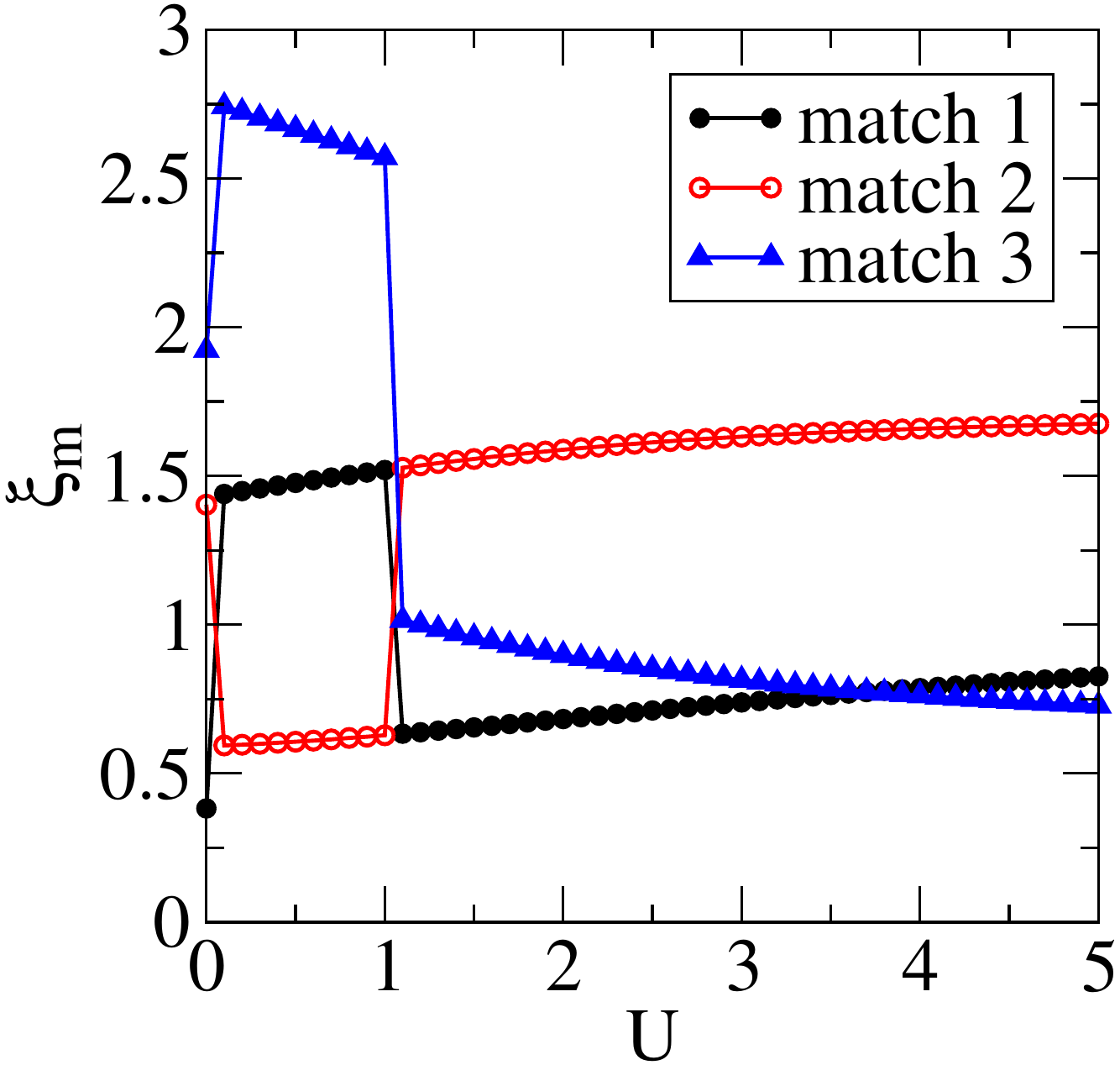}
   \end{tabular}
   \end{tabular}
   \caption{Plot of $\xi$ for each choice of $t$ and $U$ in units of the site potential 
   difference $|\epsilon_1-\epsilon_2|$.  (a) shows the range 0 to 8 for both parameters and  (b) 
expands the region 0 to 1.  The color scale is the same for both figures. 
(c) shows $\xi_m$ as a function of $U$ for three different matches when $t=1$.}
   \label{fig:map}
\end{figure}

Figure~\ref{fig:map} shows the value of $\xi$ obtained as a function of $t$ and $U$ at fixed disorder strength $|\epsilon_1 - \epsilon_2|=1$.   
Figure~\ref{fig:map}(a) covers a wide range of values in both $t$ and $U$, while Fig.~\ref{fig:map} (b) 
provides detail for values between zero and one for both parameters.
As discussed in Sec.~\ref{sec:tequals0}, in the atomic limit ($t=0$), the system is maximally localized, independent of $U$, as reflected in the figure.
The general trend illustrated in the figure is that $\xi$ increases both with increasing hopping amplitude and with increasing interaction strength.  
We find that $t$ has a stronger delocalizing effect than $U$, with $\xi$ depending on $U$ very weakly 
for $t > 2 |\epsilon_1 - \epsilon_2|$.

For almost all parameter values shown in Fig.~\ref{fig:map}(a) and (b), the optimal match (match 1) is that 
shown in Table~\ref{table:match1}. There are three exceptions.  
(i) For $t=0$ and $U>1$, the lowest energy 2-particle eigenstate is $\ket{s}$, whereas for $U<1$ it is $\ket{02}$, the state with both particles on site 2.  The optimal correspondence (match 4) switches accordingly.
(ii) For $t>0$ and $0 < U \le 1$, the optimal match (match 2) switches the sign of $\ket{w_2}$ from that in match 1.  
(iii) For $t=1$, when $U \gtrsim 4$, the optimal correspondence (match 3) is significantly rearranged from that shown in Table~\ref{table:match1}. 
Match 3 is similar to match 4 but with many signs reversed.

At $t=1$, $\xi$ has a non-monotonic variation with $U$.
To explore this in more detail, 
in Fig.~\ref{fig:map}(c) we plot $\xi_m$ values obtained from matches 1, 2 and 3.
While match 1 is optimal in the non-interacting limit, it is abruptly usurped by match 2 for nonzero $U$ up to $U=1$.  
Meanwhile match 3 gives much larger values of $\xi_m$ than the first two for small $U$ values, 
but unlike the other matches the $\xi_m$ values decline with increasing $U$ such that match 3 becomes optimal above $U \simeq 4$.

\begin{figure}[htbp] 
   \centering
   \includegraphics[width=\columnwidth]{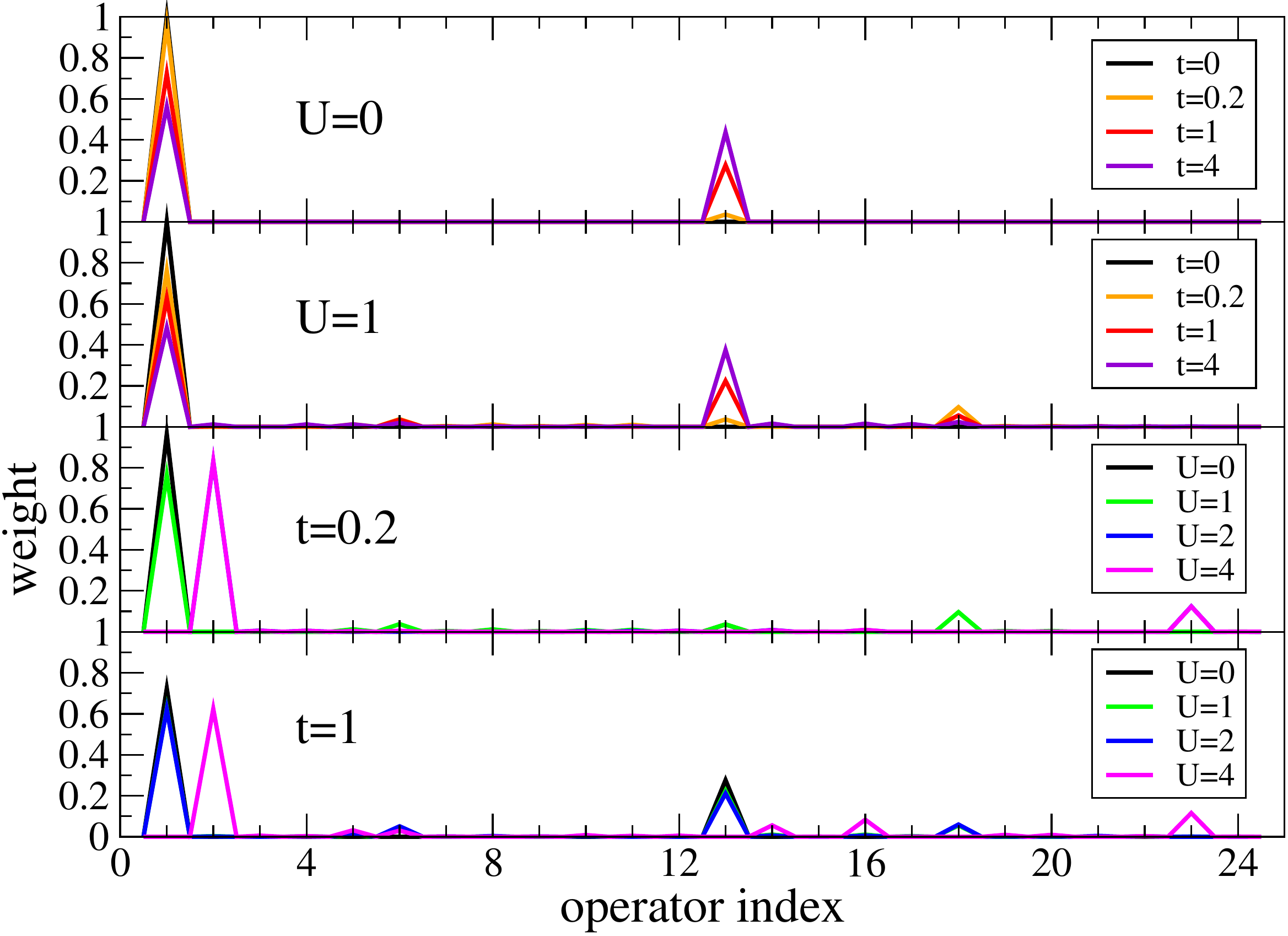} \\
   \caption{The weights of the coefficients of the normalized Fock-space creation and annihilation operators which make up $\tau_1^+$ for different parameter values.  The operator indices are as defined in Table~\ref{table:c_ops}.  }
\label{fig:tau}
\end{figure}

\begin{table}[htbp]
   \centering
\begin{tabular}{|l |r| |l |r|} \hline
1 & ${\hat c}_{1\uparrow}^{\dag}$ 	
   & 13 & ${\hat c}_{2\uparrow}^{\dag}$ \\
2 & ${\tilde {\hat n}}_{1\downarrow} {\hat c}_{1\uparrow}^{\dag}$
   & 14 & ${\tilde {\hat n}}_{2\downarrow}  {\hat c}_{2\uparrow}^{\dag}$ \\
3 & ${\tilde {\hat n}}_{2\uparrow}  {\hat c}_{1\uparrow}^{\dag}$
   & 15 & ${\tilde {\hat n}}_{1\uparrow}  {\hat c}_{2\uparrow}^{\dag}$ \\
4 & ${\tilde {\hat n}}_{2\downarrow}  {\hat c}_{1\uparrow}^{\dag}$
   & 16 & ${\tilde {\hat n}}_{1\downarrow}  {\hat c}_{2\uparrow}^{\dag}$ \\
5 & ${\hat c}_{2\downarrow}^{\dag} {\hat c}_{1\uparrow}^{\dag} {\hat c}_{1\downarrow}$ 
   & 17 & ${\hat c}_{2\downarrow}^{\dag} {\hat c}_{2\uparrow}^{\dag} {\hat c}_{1\downarrow}$  \\
6 & ${\hat c}_{1\downarrow}^{\dag} {\hat c}_{1\uparrow}^{\dag} {\hat c}_{2\downarrow}$
   & 18 & ${\hat c}_{2\uparrow}^{\dag} {\hat c}_{1\downarrow}^{\dag} {\hat c}_{2\downarrow}$ \\
7 & ${\tilde {\hat n}}_{1\downarrow} {\tilde {\hat n}}_{2\uparrow}  {\hat c}_{1\uparrow}^{\dag}$
   & 19 & ${\tilde {\hat n}}_{1\uparrow} {\tilde {\hat n}}_{1\downarrow}  {\hat c}_{2\uparrow}^{\dag}$ \\
8 & ${\tilde {\hat n}}_{1\downarrow} {\tilde {\hat n}}_{2\downarrow}  {\hat c}_{1\uparrow}^{\dag}$
   & 20 & ${\tilde {\hat n}}_{1\uparrow} {\tilde {\hat n}}_{2\downarrow}  {\hat c}_{2\uparrow}^{\dag}$ \\
9 & ${\tilde {\hat n}}_{2\uparrow}  {\tilde {\hat n}}_{2\downarrow}  {\hat c}_{1\uparrow}^{\dag}$
   & 21 & ${\tilde {\hat n}}_{1\downarrow} {\tilde {\hat n}}_{2\downarrow}  {\hat c}_{2\uparrow}^{\dag}$ \\
10 & ${\tilde {\hat n}}_{2\uparrow} {\hat c}_{2\downarrow}^{\dag} {\hat c}_{1\uparrow}^{\dag} 
          {\hat c}_{1\downarrow}$
   & 22 & ${\tilde {\hat n}}_{1\uparrow}  {\hat c}_{2\downarrow}^{\dag} {\hat c}_{2\uparrow}^{\dag} 
                 {\hat c}_{1\downarrow}$ \\
11 & ${\tilde {\hat n}}_{2\uparrow}  {\hat c}_{1\downarrow}^{\dag}{\hat c}_{1\uparrow}^{\dag}
          {\hat c}_{2\downarrow}$
   & 23 & ${\tilde {\hat n}}_{1\uparrow}  {\hat c}_{2\uparrow}^{\dag} {\hat c}_{1\downarrow}^{\dag} 
                 {\hat c}_{2\downarrow}$ \\
12 & ${\tilde {\hat n}}_{1\downarrow} {\tilde {\hat n}}_{2\uparrow} 
          {\tilde {\hat n}}_{2\downarrow}  {\hat c}_{1\uparrow}^{\dag}$
   & 24 & ${\tilde {\hat n}}_{1\uparrow} {\tilde {\hat n}}_{1\downarrow} 
                 {\tilde {\hat n}}_{2\downarrow}  {\hat c}_{2\uparrow}^{\dag}$ \\ \hline
\end{tabular}
\caption{Fock space operators contributing to $\tau_1^+$ and their indices.  Weights are shown in Fig.~\ref{fig:tau}.  
Note that ${\tilde {\hat n}}_i \equiv {\hat n}_i-\frac{1}{2}{\hat I}$.}
   \label{table:c_ops}
\end{table}

In addition to studying the evolution of $\xi$ we investigated the nature of the resulting $\tau_i^+$ operators.
We illustrate this with a detailed decomposition of the operator $\tau_1^+$.
In the parameter range we have considered, only 24 of the 256 possible combinations of $c$ operators contribute to $\tau_1^+$, and these are listed in Table~\ref{table:c_ops}.
Figure~\ref{fig:tau} shows the weight of the contribution each of these (normalized) operators makes to $\tau_1^+$ for a range of different parameters.  

In the top panel of Fig.~\ref{fig:tau}, the black line corresponds to $U=0$ and $t=0$.  
The first operator has weight one and no others contribute, 
corresponding to $\tau_1^+={\hat c}_{1\uparrow}^{\dag}$ as discussed in Sec.~\ref{sec:tequals0}.  
As $t$ is increased for $U = 0$, weight shifts from ${\hat c}_{1\uparrow}^{\dag}$ (operator 1) 
to ${\hat c}_{2\uparrow}^{\dag}$ (operator 13) consistent with the discussion in Sec.~\ref{sec:Uequals0}.
The second panel of Fig.~\ref{fig:tau} shows the same sequence of $t$ values with $U=1$.  
The $t=0$ configuration is identical to the non-interacting case, and even with hopping, while other operators begin to contribute, ${\hat c}_{1\uparrow}^{\dag}$ and ${\hat c}_{2\uparrow}^{\dag}$ remain dominant.
The third and fourth panels of Fig.~\ref{fig:tau} show instead a range of $U$ values at fixed $t$.  
For low $t$ and $U$ values, these maximally localized LIOMs have strong overlap with single-particle occupation numbers obtained in the absence of interactions, a result also noted in other systems.\cite{Rademaker2016arxiv}
For higher $U$ values, when the optimal match switches to match 3, there is a qualitative change as ${\tilde {\hat n}}_{1\downarrow} {\hat c}_{1\uparrow}^{\dag}$ (operator 2) replaces ${\hat c}_{1\uparrow}^{\dag}$ as the primary contribution.
Not shown, when $t$ is large, independent of $U$, the structure is similar to the atomic limit, with roughly equal weight on 
${\hat c}_{1\uparrow}^{\dag}$ and  ${\hat c}_{2\uparrow}^{\dag}$.

\begin{figure}[htbp] 
   \centering
   \includegraphics[width=\columnwidth]{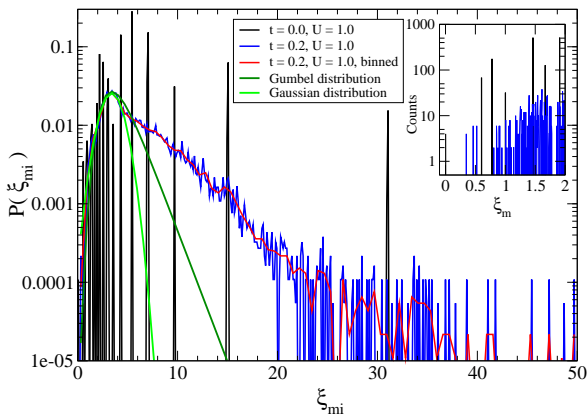} \\
   \caption{The distribution of $\xi_{mi}$ values over all the different matches for $t=0.2$, $U=1.0$ and the atomic limit, $t=0.0$, $U=1.0$. Also shown are Gumbel and Gaussian fits to the distribution of $\xi_{mi}$ values for $t=0.2$, $U=1.0$. The inset shows the counts of $\xi_m$ for small $\xi_m$ for the same two sets of parameter values.}
   \label{fig:distribution}
\end{figure}

The $\tau_1^+$ operators shown in Fig.~\ref{fig:tau} are the most local by the criterion that for these matches, $\xi_m = \xi$.  
This most-localized choice is highly non-generic.
Figure~\ref{fig:distribution} illustrates this point by showing 
the distribution of $\xi_{mi}$ values for two different parameter sets:  $(t,U)=(0,1.0)$ and $(t,U)=(0.2,1.0)$.
When either $U$ or $t$ is zero, the distribution has sharp peaks separated by wide gaps.
While the distribution is always discrete when viewed with perfect resolution, when $U$ and $t$ are both non-zero, 
it is more convenient to view the binned distribution. The inset shows the low end of the distribution of $\xi_{m}$ values, highlighting the point that the most localized match is usually considerably more localized than
 the next best match.

We note that the peaks of these distributions are at considerably larger values than the optimum, and 
that they have very long tails at large values of $\xi_{mi}$.  
We show in Fig.~\ref{fig:distribution} a comparison of the distribution of $\xi_{mi}$ values to a Gaussian and a Gumbel distribution, 
for $t=0.2$, $U=1.0$, neither of which is a particularly good fit for large values of $\xi_{mi}$, although the 
Gumbel distribution is certainly a better fit, and fits the data over a large range of $\xi_{mi}$ values for
larger $t$.  The strength of the tail of the distribution at large $\xi_{mi}$ is particularly notable,
since even the Gumbel distribution, which arises in extreme value statistics, is insufficient to capture the tail.

\section{Discussion}
\label{sec:discussion}

In this work we considered the two-site Anderson-Hubbard model as a toy model of interacting electrons in the presence of disorder.  
We constructed LIOMs for this model, and searched through possible matches between energy eigenstates and 
$\tau$ basis states to obtain the most local choice of LIOMs as defined by our localization measure, Eq.~(\ref{loc_meas}).
We mapped the degree of localization of this most local choice as a function of parameters 
$t$ and $U$, noting the variation not only in the localization measure but also in the optimal match.
We explicitly presented the composition of the optimal LIOMs in terms of standard creation/annihilation operators, 
and explored the distribution of locality obtained in all matches in our search.

Since the observation that properties of fully many-body localized systems could be described in terms of conserved pseudospins, numerous schemes have been suggested to identify LIOM in particular systems.  
The use of a direct unitary transformation to map from the energy eigenbasis to a basis of local operators has been raised but rarely 
implemented.\cite{Serbyn2013,Huse2014,Chandran2015,Pekker2016,He2016,Wahl2016}
Other proposals include renormalization\cite{Vosk2013,Pekker2016,Monthus2016} and perturbative 
approaches\cite{Imbrie2016prl,Imbrie2016jsp,Ros2015} as well as a displacement operator method\cite{Rademaker2016,Rademaker2016arxiv}, minimizing 
the commutator of approximate LIOM with the Hamiltonian\cite{He2016,OBrien2016} and identifying ``natural orbitals'' from the one 
particle density matrix.\cite{Bera2015,Bera2016} 
Concurrent with the effort to identify LIOMs is exploration of their optimization.
While Rademaker {\it et al}.\cite{Rademaker2016arxiv} recently suggested that one may wish to optimize from alternative perspectives (e.g. information theory might suggest minimizing the {\em number} of couplings between pseudospins), the primary focus has been on identifying a localization length which diverges at the MBL transition, analogous to the localization length defined by the size of single-particle states in Anderson localization.
The most direct analogy is to measure the size of the LIOMs themselves.
Examples include the IPR,\cite{Bera2015} the support,\cite{Chandran2015,OBrien2016} 
and the overlap with local operators in Fock space.\cite{Rademaker2016arxiv}
Other work has focused on the length associated with the interactions between pseudospins ($\beta_{ij}$, $\gamma_{ijk}$, \ldots in Eq.~(\ref{eq:lbit_Hamiltonian})).\cite{Huse2014,Chandran2015,Rademaker2016,Pekker2016} 
As shown by Rademaker and Ortu\~{n}o,\cite{Rademaker2016} the lengthscales for pseudospins and interaction parameters can be quite different, even when both exhibit exponential decay with distance.  

Two advantages of the unitary mapping approach to identifying integrals of motion used here are (i) that it is effective and equally 
efficient independent of the level of localization in the system, and (ii) that it makes transparent the 
fact that choice is involved in identifying LIOM.  
Perturbative schemes break down as the transition is approached and the displacement operator method, 
while still applicable, requires increasing numbers of transformations.  
Moreover, these methods focus on a systematic progression towards a single set 
of LIOMs and do not provide perspective on the discarded alternatives.
Clearly there are also disadvantages to our approach, first among them being the necessity to work with small 
systems, not only to have access to a complete set of eigenstates but also to make the search through options manageable.
A general challenge for any scheme trying to optimize the locality of LIOMs is the choice of measure
of localization to use to discriminate between matches. The measure (Eq.~(\ref{loc_meas})) that we used
focuses on the extent of the LIOMs themselves as opposed to the range of their interactions 
and aims to treat delocalization in both a single-particle and many-body sense on the same footing.
We hope that the strengths of the unitary mapping approach may provide insights that can inform the development of more widely applicable methods for optimizing LIOMs.

Although we had access in this work to an analytic solution of the model, the same procedure for identifying LIOM can be used with numerically obtained eigenstates, so long as all eigenstates and not just the ground state are known.  
The feature of our work which would be more difficult without an analytic solution is the tracking of specific matches.  A numerical algorithm may not consistently assign the same sign to a particular eigenvector as parameters are varied.  This will not effect the optimal value of $\xi$ obtained for any given parameter set, provided the search through matches allows for all sign options.  However, identifying whether the chosen match is the same or different from matches at other parameter values may be more difficult.

The importance of the Hubbard model as a minimal model of strongly correlated electrons motivates extending our work to larger system sizes using numerical or recently developed tensor network methods.\cite{Wahl2016} In addition, recent work\cite{Znidaric2016}
has suggested that only charge, and not spin, localizes in the disordered Hubbard chains, 
so that MBL in the Anderson-Hubbard model may have a different character than in other commonly studied MBL systems.  
The addition of a random magnetic field can lead to localization of spin as well as charge degrees of freedom\cite{Mondaini2015,Znidaric2016}
and should be explored further.

In summary, the work presented here provides researchers in strongly correlated electrons with a 
concrete picture of the form of the LIOMs in a Hubbard system, and it provides the many-body localization community with an exploration of the full range of options for LIOMs in a toy system, offering perspective on the evolution of the nature of the pseudospin operators away from the fully localized limit.

\acknowledgements
We gratefully acknowledge support by the Natural Sciences and Engineering Research Council (NSERC) of Canada.  
We thank David Huse and Louk Rademaker for helpful discussions.

\begin{appendix}

\section{Eigenvalues and eigenvectors of the two-site Anderson-Hubbard model}
\label{app:2siteAHM}
In this appendix we summarize the analytic expressions for the eigenvalues and eigenvectors
of the two-site Anderson-Hubbard model.  We first list the basis which we 
use to express the eigenstates -- this is a slightly modified Fock basis in which the 
states $\ket{\uparrow\downarrow}$ and $\ket{\downarrow\uparrow}$ are combined to form 
spin singlet and triplet states.

\begin{table}[ht]
       \begin{tabular}{lcl}
               $\ket{00}$ &=& $\ket{00}$ \\ \hline
               $\ket{\uparrow 0} $ &=& $c_{1\uparrow}^{\dag}\ket{00}$ \\
               $\ket{0 \uparrow} $ &=& $c_{2\uparrow}^{\dag}\ket{00}$ \\
               $\ket{\downarrow 0}$ &=& $c_{1\downarrow}^{\dag}\ket{00} $ \\
               $\ket{0 \downarrow} $ &=& $c_{2\downarrow}^{\dag} \ket{00} $ \\ \hline
               $\ket{\uparrow \uparrow}$ &=& $c_{2\uparrow}^{\dag} c_{1\uparrow}^{\dag} \ket{00} $ \\
               $\ket{t_0}$
               &=& ${1 \over \sqrt{2}} \left( c_{2\downarrow}^{\dag} c_{1\uparrow}^{\dag} + c_{2\uparrow}^{\dag} 
                  c_{1\downarrow}^{\dag} \right) \ket{00}$ \\
                 $\ket{\downarrow \downarrow}$ &=& c$_{2\downarrow}^{\dag} c_{1\downarrow}^{\dag} \ket{00}$ \\ \hline
               $\ket{20}$ &=& $c_{1\downarrow}^{\dag} c_{1\uparrow}^{\dag} \ket{00}$ \\
              $\ket{s}$   &=& ${1 \over \sqrt{2}} \left( c_{2\downarrow}^{\dag} c_{1\uparrow}^{\dag} 
			    - c_{2\uparrow}^{\dag} c_{1\downarrow}^{\dag} \right) \ket{00} $ \\
	      $\ket{02}$ &=& $c_{2\downarrow}^{\dag} c_{2\uparrow}^{\dag} \ket{00} $ \\ \hline
	      $\ket{2 \uparrow}$ &=& $c_{2 \uparrow}^{\dag} c_{1\downarrow}^{\dag} c_{1\uparrow}^{\dag} \ket{00}$ \\
	      $\ket{\uparrow 2}$ &=& $c_{2 \downarrow}^{\dag} c_{2\uparrow}^{\dag} c_{1\uparrow}^{\dag} \ket{00}$ \\
	      $\ket{2 \downarrow}$ &=& $c_{2 \downarrow}^{\dag} c_{1\downarrow}^{\dag} c_{1\uparrow}^{\dag} \ket{00}$ \\
	      $\ket{\downarrow 2}$ &=& $c_{2 \downarrow}^{\dag} c_{2\uparrow}^{\dag} c_{1\downarrow}^{\dag} \ket{00}$ \\ \hline
	      $\ket{22}$ &=& $c_{2\downarrow}^{\dag} c_{2\uparrow}^{\dag} c_{1\downarrow}^{\dag} c_{1\uparrow}^{\dag}\ket{00} $
	\end{tabular}
   \caption{Modified Fock basis}
 \label{table:fock}
\end{table}

\begin{table}[ht]
	\begin{tabular}{|l |l |l |} \hline
		State & Eigenvalue & Eigenstate \\ \hline
		$0$ & $0$ & $ \ket{00} $ \\ \hline
		$p\uparrow$ & $E_{1p} =  x + \sqrt{y^2 + t^2}$ & $\alpha\ket{\uparrow 0} - \beta \ket{0 \uparrow} $ \\
		$m\uparrow$ & $E_{1m} =  x - \sqrt{y^2 + t^2}$ & $ \beta\ket{\uparrow 0} + \alpha \ket{0 \uparrow}$ \\
		$p\downarrow$ & $E_{1p}$ & $ \alpha \ket{\downarrow 0} -\beta \ket{0 \downarrow} $ \\
		$m\downarrow$ & $E_{1m}$ & $ \beta \ket{\downarrow 0} +  \alpha \ket{0 \downarrow} $ \\ \hline
		$t\uparrow$ & $E_{2t} = 2x$ & $\ket{\uparrow\uparrow}$ \\
		$t_0$ & $E_{2t}$ & $\ket{t_0}$ \\
		$t\downarrow$ & $E_{2t}$ & $\ket{\downarrow\downarrow}$ \\ \hline
		$w_1$ & $E_{w_1}$
		& $ \phi_{1\ell}\ket{2 0} +  \phi_{1s}\ket{s} +  \phi_{1r}\ket{0 2} $ \\
		$w_2$ & $E_{w_2}$
		& $ \phi_{2\ell}\ket{2 0} +  \phi_{2s}\ket{s} +  \phi_{2r}\ket{0 2} $ \\
		$w_3$ & $E_{w_3}$
		& $  \phi_{3\ell} \ket{2 0} +  \phi_{3s} \ket{s} + \phi_{3r} \ket{0 2}  $ \\ \hline
		$3p\uparrow$ & $E_{3p} = E_{1p} + 2x + U $
		& $\alpha \ket{2\uparrow} + \beta \ket{\uparrow 2}$ \\
		$3m\uparrow$ & $E_{3m} = E_{1m} + 2x + U $
		& $ -\beta\ket{2\uparrow} + \alpha\ket{\uparrow 2}$ \\
		$3p\downarrow$ & $E_{3p} $
		& $ \alpha\ket{2\downarrow} + \beta\ket{\downarrow 2}$ \\
		$3m\downarrow$ & $E_{3m} $
		& $ -\beta\ket{2\downarrow} +  \alpha\ket{\downarrow 2}$ \\ \hline
		$4$ & $ E_4 = 4x + 2U $
		& $ \ket{22} $ \\ \hline
	\end{tabular}
	\caption{Table of eigenvalues and eigenvectors for the two-site Anderson-Hubbard model}
	\label{table:evef}
\end{table}
The eigenvalues and eigenvectors of the two-site Anderson-Hubbard model are summarized in Table\ \ref{table:evef},
where we defined
\begin{equation} 
	x = \frac{\epsilon_1 + \epsilon_2}{2}, \quad \quad y = \frac{\epsilon_1 - \epsilon_2}{2}.
\end{equation}
    [Note that these definitions differ by a factor of $\sqrt{2}$ from the definition of $x$ and $y$
      in Ref.~\onlinecite{Johri2012}.]  The coefficients $\alpha$ and $\beta$ are given by:
\begin{equation} 
\left(\begin{array}{c} \alpha \\ \beta \end{array}\right) 
= \frac{1}{\sqrt{t^2 + \left( y + \sqrt{y^2 + t^2}\right)^2}}  \left(\begin{array}{c} y + \sqrt{y^2 + t^2} \\ t 
\end{array}\right) .
\end{equation}

 The eigenvalues for the $\ket{w_1}$, $\ket{w_2}$ and $\ket{w_3}$ states are
\begin{equation}
E_{w_i} = 2x + \frac{2U}{3} - \frac{2}{3} \sqrt{U^2 + 12 (y^2 + t^2)} \cos\left(\frac{\theta_i}{3}\right),
\end{equation}
 where
 \begin{widetext}
\begin{equation} 
\theta_i = \left\{\begin{array}{cc} \theta + 2\pi, & i = 1, \\ \theta - 2\pi, & i = 2, \\ \theta, & i = 3 
 \end{array} \right. ,\quad \quad \quad 
  \theta = \cos^{-1} \left[\frac{U(U^2 - 36y^2 + 18t^2)}{\left(U^2 + 12y^2 + 12t^2\right)^\frac{3}{2}}\right].
\end{equation}
 The coefficients in the expressions for the $\ket{w_1}$, $\ket{w_2}$ and $\ket{w_3}$ states in terms of the
 modified Fock basis are:
\begin{equation}
\left(\begin{array}{c} \phi_{i,l} \\ \phi_{i,s} \\ \phi_{i,r} \end{array} \right) 
   =  \frac{{\rm sgn}(4y^2 - B_i^2)}{\sqrt{\left(4y^2 - B_i^2\right)^2 + 4t^2\left(4y^2 + B_i^2\right)}} 
   \left(\begin{array}{c} \sqrt{2} t(2y - B_i) \\ \left(4y^2 - B_i^2\right) \\ - \sqrt{2}t(2y+B_i) \end{array}\right),
\end{equation}
  where
\begin{equation}
B_i = \frac{U}{3} + \frac{2}{3} \sqrt{U^2 + 12 (y^2 + t^2)} \cos\left(\frac{\theta_i}{3}\right).
\end{equation}

\section{Pseudospin Hamiltonian}
\label{app:H_tauz}

Our focus has been on the nature of the pseudospin operators themselves.
It is nonetheless worth clarifying that having identified the $\tau_i^+$ operators as described,
writing the corresponding pseudospin Hamiltonian is straightforward.
Any Hamiltonian may be written 
\begin{eqnarray}
{\cal H} &=& \sum_\alpha E_{\alpha} {\hat d}_{\alpha}^{\dag} {\hat d}_{\alpha} ,
\label{H_eigen}
\end{eqnarray}
where ${\hat d}_{\alpha}^{\dag}$ creates the eigenstate $\alpha$ with energy $E_{\alpha}$.
Once we have constructed our correspondence table, 
we then know all the eigenstate creation operators in terms of $\tau_i^+$:
${\hat d}_1^{\dag}$ is the identity operator,  ${\hat d}_2^{\dag}=\tau_1^+$, etc, through to ${\hat d}_{16}^{\dag}=-\tau_4^+ \tau_3^+ \tau_2^+ \tau_1^+$ in this case.
Putting these expressions into Eq. (\ref{H_eigen}) and using $\tau_i^z = \tau_i^+ \tau_i^- - {1 \over 2}$, the pseudospin Hamiltonian is rapidly generated.
For the specific correspondence shown in Table~\ref{table:match1} the expression is

\begin{eqnarray}
{\cal H} &=&  \left[ E_{1p} + {1 \over 2}(2E_{t} +  E_{w1}) + {1 \over 4}(2E_{3p} + E_{3m}) + {1 \over 8}E_4 \right] \tau_1^z 
+  \left[ E_{1m} + {1 \over 2}(E_{t} + E_{w2} + E_{w3}) + {1 \over 4}(E_{3p}+2E_{3m}) + {1 \over 8} E_4 \right] \tau_2^z 	\nonumber \\ 
  & & + \left[ E_{1p} + {1 \over 2}(E_{t} + E_{w1} + E_{w2}) + {1 \over 4}(2E_{3p}+E_{3m}) + {1 \over 8} E_4 \right] \tau_3^z 
+ \left[ E_{1m} + {1 \over 2}(2E_{t} + E_{w3}) + {1 \over 4}(E_{3p}+2E_{3m}) + {1 \over 8} E_4 \right] \tau_4^z \nonumber \\ 
  & & +  \left[ E_{t} + {1 \over 2}(E_{3p}+E_{3m})+{1 \over 4} E_4 \right]\tau_2^z \tau_1^z  
	+  \left[ E_{w1} + {1 \over 2}(2E_{3p})+{1 \over 4} E_4 \right] \tau_3^z \tau_1^z 
+  \left[ E_{t} + {1 \over 2}(E_{3p}+E_{3m})+{1 \over 4} E_4 \right] \tau_4^z \tau_1^z   \nonumber \\
   & & 	+  \left[ E_{w2} + {1 \over 2}(E_{3p}+E_{3m})+{1 \over 4} E_4 \right] \tau_3^z \tau_2^z 
+  \left[ E_{w3} + {1 \over 2}(2E_{3m})+{1 \over 4} E_4 \right]  \tau_4^z \tau_2^z 
	+ \left[ E_{t} + {1 \over 2}(E_{3p}+E_{3m})+{1 \over 4} E_4 \right] \tau_4^z \tau_3^z \nonumber \\
	& & + \left[ E_{3p} + {1 \over 2}E_4 \right] \tau_3^z \tau_2^z \tau_1^z 
	+  \left[ E_{3m} + {1 \over 2}E_4 \right] \tau_4^z \tau_2^z \tau_1^z 
	+  \left[ E_{3p} + {1 \over 2}E_4 \right] \tau_4^z \tau_3^z \tau_1^z 
	+  \left[ E_{3m} + {1 \over 2}E_4 \right] \tau_4^z \tau_3^z \tau_2^z \nonumber \\
& & + E_4 \tau_4^z \tau_3^z \tau_2^z \tau_1^z  + {\rm constant}  .
\end{eqnarray}
\end{widetext}

\end{appendix}

\end{document}